\documentclass[twocolumn, aps, prb]{revtex4}
\usepackage{graphicx}
\usepackage{amssymb}
\usepackage{amsmath}
\newcommand{\GVec}[1]{\mbox{\boldmath$#1$}}
\def\Journal #1,#2,#3,#4#5#6#7{#1 {\bf #2}, #3 (#4#5#6#7)}
\begin{document}
%
\title{Anomalous orbital magnetism in Dirac-electron systems:\\
Role of pseudo-spin paramagnetism}
\author{Mikito Koshino and Tsuneya Ando}
\affiliation{
Department of Physics, Tokyo Institute of Technology,
2--12--1 Ookayama, Meguro-ku, Tokyo 152-8551, Japan}
\date{\today}
%
\begin{abstract}
The orbital diamagnetic susceptibility is calculated in monolayer and bilayer graphenes with band gap as well as in three-dimensional Dirac systems.
It is demonstrated that the pseudo-spin degree of freedom such as valleys produces paramagnetic susceptibility in an equal manner as the real spin dominating over the Landau diamagnetism. 
The pseudo-spin paramagnetism explains the origin of a singular diamagnetism which is present only in the band-gap region and disappears rapidly inside the conduction and valence bands.
\end{abstract}
%
\maketitle
%
\section{Introduction}
\label{sec_intr}
%
The magnetism of conventional metal 
is composed of two different contributions,
the spin component known as the Pauli paramagnetism and the orbital component
as the Landau diamagnetism.
In condensed matter systems, the orbital magnetism
sensitively depends on the detail of the electronic band structure,
and sometimes largely deviates from the Landau diamagnetism.
Particularly, narrow gap materials such as 
graphite\cite{McClure_1956a,McClure_1960a,Sharma_et_al_1974a} or bismuth \cite{Wolff_1964a,Fukuyama_and_Kubo_1969a,Fukuyama_and_Kubo_1970a}
exhibit a singular behavior in the orbital susceptibility
near the energy gap.
In this paper, we show that 
the anomalous orbital magnetism in narrow gap systems
can be understood in terms of pseudo-spin paramagnetism,
which arises from the extra degree of freedom in the
orbital motion of electrons.
\par
%
Graphene monolayer recently fabricated\cite{Novoselov_et_al_2004a,Novoselov_et_al_2005a,Zhang_et_al_2005a} is a zero-gap system in which the conduction and valance bands stick together at $K$ and $K'$ points located at inequivalent corners of the Brillouin zone, called valleys.\cite{McClure_1956a,Slonczewski_and_Weiss_1958a,DiVincenzo_and_Mele_1984a,Semenoff_1984a,Shon_and_Ando_1998a,Zheng_and_Ando_2002a,Ando_2005a,Gusynin_and_Sharapov_2005a,Peres_et_al_2006a}
The system is characterized by chiral quasiparticles with opposite chirality in each valley and a linear dispersion reminiscent of massless Dirac fermions.
At the Dirac point where two bands cross each other, the magnetic susceptibility has a singularity expressed as a delta function in Fermi energy $\varepsilon_F$ and disappears otherwise.\cite{McClure_1956a,Sharapov_et_al_2004a,Fukuyama_2007a,Nakamura_2007a,Koshino_and_Ando_2007b,Ghosal_et_al_2007a,Ando_2007d,Koshino_et_al_2009a}
\par
%
Bilayer graphene composed of a pair of graphene layers\cite{Novoselov_et_al_2006a,Ohta_et_al_2006a,Castro_et_al_2007b,Oostinga_et_al_2008a} has a zero-gap structure with quadratic dispersion,\cite{McCann_and_Falko_2006a,Guinea_et_al_2006a,Lu_et_al_2006a,Lu_et_al_2006b,McCann_2006a,Koshino_and_Ando_2006a,Nilsson_et_al_2006b,Partoens_and_Peeters_2006a,Partoens_and_Peeters_2007a} leading to a less singular, logarithmic peak of the susceptibility.\cite{Safran_1984a,Koshino_and_Ando_2007c}
The orbital magnetism was also studied for related materials, such as graphite intercalation compounds,\cite{Safran_and_DiSalvo_1979a,Safran_1984a,Blinowski_and_Rigaux_1984a,Saito_and_Kamimura_1986a} carbon nanotube,\cite{Ajiki_and_Ando_1993a,Ajiki_and_Ando_1993b,Ajiki_and_Ando_1995c,Yamamoto_et_al_2008a} few-layer graphenes,\cite{Koshino_and_Ando_2007c,Nakamura_and_Hirasawa_2008a,Castro_Neto_et_al_2009a} and organic compounds having Dirac-like spectrum.\cite{Kobayashi_et_al_2008a}
In both monolayer\cite{Semenoff_1984a,Ludwig_et_al_1994a} and bilayer graphenes,\cite{Ohta_et_al_2006a,Castro_et_al_2007b,Oostinga_et_al_2008a,McCann_and_Falko_2006a,Lu_et_al_2006a,Lu_et_al_2006b,Guinea_et_al_2006a,McCann_2006a,Nilsson_et_al_2006b,Ando_and_Koshino_2009a,Ando_and_Koshino_2009b} certain asymmetric potential opens an energy gap at the band touching point.
\par
%
In this paper, we calculate the orbital magnetism of several Dirac-like systems with gap and show that the pseudo-spin degree of freedom such as valleys in graphene produces paramagnetism in the same manner as the real spin and gives an essential contribution to the singular diamagnetic behavior.
In Sec.\ \ref{sec_mono}, we calculate the susceptibility of the monolayer graphene with varying gap.
The singular susceptibility-change in varying $\varepsilon_F$ is understood in terms of valley-induced paramagnetism.
We extend the analysis to the bilayer graphene in Sec.\ \ref{sec_bi} and to a three-dimensional Dirac system corresponding to bismuth\cite{Wolff_1964a,Fukuyama_and_Kubo_1969a,Fukuyama_and_Kubo_1970a,Fuseya_et_al_2009a} in Sec.\ \ref{sec_3d}.
A brief conclusion is presented in Sec.\ \ref{sec_conc}.
\par
%
\section{Monolayer graphene}
\label{sec_mono}
%
Graphene is composed of a honeycomb network of carbon atoms, where a unit cell contains a pair of sublattices, denoted by $A$ and $B$.
Electronic states in the vicinity of $K$ and $K'$ points in the Brillouin zone are well described by the effective mass approximation.\cite{McClure_1956a,Slonczewski_and_Weiss_1958a,DiVincenzo_and_Mele_1984a,Semenoff_1984a,Ando_2005a,Shon_and_Ando_1998a,Zheng_and_Ando_2002a,Gusynin_and_Sharapov_2005a,Peres_et_al_2006a}
Let $|A\rangle$ and $|B\rangle$ be the Bloch functions at the $K$ point, corresponding to the $A$ and $B$ sublattices, respectively.
In a basis $(|A\rangle,|B\rangle)$, the Hamiltonian for the monolayer graphene around the $K$ point becomes\cite{McClure_1956a,Slonczewski_and_Weiss_1958a,DiVincenzo_and_Mele_1984a,Semenoff_1984a,Ando_2005a,Shon_and_Ando_1998a,Zheng_and_Ando_2002a,Gusynin_and_Sharapov_2005a,Peres_et_al_2006a}
%
\begin{equation}
{\mathcal H}^K = \begin{pmatrix} \Delta & v \pi_- \\ v \pi_+ & -\Delta \end{pmatrix} ,
\label{eq:H_mono}
\end{equation}
%
where $v$ is the velocity, $\pi_\pm = \pi_x \pm i \pi_y$, and $\GVec{\pi} = -i\hbar \GVec{\nabla} + (e/c) {\bf A}$ with vector potential ${\bf A}$ giving external magnetic field ${\bf B} = \GVec{\nabla}\times {\bf A}$.
In the following, we shall completely neglect the spin Zeeman energy because the spin splitting is much smaller than Landau-level separations.
The Hamiltonian at the $K'$ point is obtained by exchanging $\pi_\pm$ in Eq.\ (\ref{eq:H_mono}).
\par
%
The diagonal terms $\pm\Delta$ represent the potential asymmetry between $A$ and $B$ sites, which opens an energy gap at the Dirac point.\cite{Semenoff_1984a,Ludwig_et_al_1994a}
This can arise when graphene is placed on a certain substrate material.
In fact, band-gap opening is observed in graphene epitaxially grown on a SiC substrate.\cite{Zhou_et_al_2007a,Zhou_et_al_2008b}
From a theoretical point of view, the singular behavior in ideal graphene with vanishing gap is better understood by taking the limit $\Delta\rightarrow0$, as will be shown below.
We can safely assume $\Delta\ge0$ without loss of generality.
\par
%
The energy band at $B=0$ is given by 
%
\begin{eqnarray}
 \varepsilon_{s}(p) = s \sqrt{v^2p^2 + \Delta^2},
\quad (s=\pm1)
\end{eqnarray}
%
with electron momentum ${\bf p} = (p_x,p_y)$ and $p=\sqrt{p_x^2+p_y^2}$.
The density of states is \cite{Ludwig_et_al_1994a}
%
\begin{eqnarray}
D(\varepsilon) = \frac{g_v g_s |\varepsilon|}{2\pi\hbar^2 v^2} \theta (|\varepsilon|-|\Delta|),
\label{eq:DensityOfStatesInMonolayerGraphene}
\end{eqnarray}
%
where $g_s=2$ and $g_v=2$ represent the degrees of freedom associated with spin and valley, respectively, and $\theta(t)$ is a step function, defined by
%
\begin{equation}
\theta(t)= \left\{ \begin{array}{cc} 1 & (t>0); \\ 0 & (t<0). \end{array} \right.
\end{equation}
%
\par
%
The Landau-level spectrum can be found using the relation $\pi_+=(\sqrt{2}\hbar/l_B) a^\dagger$ and $\pi_-=(\sqrt{2}\hbar/l_B)a$, where $l_B = \sqrt{c\hbar/(eB)}$ is magnetic length and $a^\dagger$ and $a$ are raising and lowering operators for usual Landau-level wave functions, respectively.
The eigenenergy at $K$ point becomes
%
\begin{eqnarray}
\varepsilon_{n}^{K} = {\rm sgn}_-(n) \sqrt{(\hbar\omega_B)^2 |n| + \Delta^2} \quad (n=0,\pm1,\pm2,\cdots) ,
\label{eq:mono_landauK}
\end{eqnarray}
%
where $\hbar \omega _B = \sqrt{2}\hbar v/l_B$ and
%
\begin{equation}
{\rm sgn}_\pm(n) = \left\{ \begin{array}{cc} +1 & (n>0); \\ \pm1 & (n=0); \\ -1 & (n<0). \end{array} \right.
\end{equation}
%
The corresponding wavefunction is
%
\begin{eqnarray}
\Phi_{n}^{K} = \begin{pmatrix} \sin(\alpha_n/2)\, \phi_{|n|-1} \\ \cos(\alpha_n/2)\, \phi_{|n|} \end{pmatrix} ,
\label{eq:mono_waveK}
\end{eqnarray}
%
where $\alpha_n$ satisfies
%
\begin{eqnarray}
&& \sin \alpha_n = {\hbar\omega_B \sqrt{|n|} \, {\rm sgn}_-(n) \over \sqrt{(\hbar\omega_B)^2 |n| + \Delta^2} } , \\
&& \cos \alpha_n = - {\Delta \, {\rm sgn}_-(n) \over \sqrt{(\hbar\omega_B)^2 |n| + \Delta^2} } ,
\end{eqnarray}
%
and $\phi_{n}$ is the usual Landau-level wave function, where $\phi_n$ with $n<0$ should be regarded as 0.
\par
%
The Landau level $n=0$ lies just at the top of the valence band, i.e., $\varepsilon_0^K=-\Delta$ because $\alpha_0=0$, and its amplitude is only at the $B$ site.
Similarly, for Landau levels lying in the vicinity of the valence-band top, i.e., $n\le0$ satisfying $\hbar\omega_B\sqrt{|n|}\ll\Delta$, we have $\alpha_n\approx0$, showing that the amplitude of the wave function is significant only at the B site.
For those in the conduction band $n>0$, on the other hand, we have $\alpha_n\approx\pi$, showing that the amplitude is significant only at the $A$ site.
\par
%
For the $K'$ point, on the other hand, the eigenenergy is given by Eq.\ (\ref{eq:mono_landauK}) with ${\rm sgn}_-(n)$ being replaced by ${\rm sgn}_+(n)$ and the eigenfunction is given by
%
\begin{eqnarray}
\Phi_{n}^{K'} = \begin{pmatrix} \sin(\alpha'_n/2)\, \phi_{|n|} \\ \cos(\alpha'_n/2)\, \phi_{|n|-1} \end{pmatrix} ,
\label{eq:mono_waveK'}
\end{eqnarray}
%
where $\alpha'_n$ is obtained from $\alpha_n$ by replacing ${\rm sgn}_-(n)$ by ${\rm sgn}_+(n)$.
The Landau level $n=0$ lies just at the bottom of the conduction band, i.e., $\varepsilon_0^{K'}=+\Delta$ because $\alpha'_0=\pi$, and its amplitude is only at the $A$ site.
Similarly, for low-lying Landau levels in the conduction band $n\ge0$, we have $\alpha'_n\approx\pi$, showing that the amplitude of the wave function is significant only at the $A$ site.
For those in the valence band $n<0$, on the other hand, we have $\alpha'_n\approx0$, showing that the amplitude is significant only at the $B$ site.
\par
%
The Landau levels of $n \ne 0$ are doubly degenerate between the $K$ and $K'$ valleys, while those of $n=0$ are not. 
Therefore, by defining
%
\begin{eqnarray}
&& \varepsilon_s(x_n) = s \sqrt{x_n + \Delta^2} , \\
&& x_n = (\hbar \omega_B)^2 n 
\label{eq:x}
\end{eqnarray}
%
the thermodynamical potential at temperature $T$ then becomes
%
\begin{eqnarray}
\Omega = -\frac{1}{\beta} \frac{g_vg_s}{2\pi l_B^2} \sum_s
\sum_{n=0}^\infty 
\varphi\big[\varepsilon_s(x_n) \big] 
\Big(1-\frac{\delta_{n0}}{2}\Big), 
\label{eq:omega}
\end{eqnarray}
%
where $\beta = 1/k_B T$ and
%
\begin{equation}
\varphi(\varepsilon) = \log \big[ 1 + e^{-\beta(\varepsilon - \zeta)} \big] ,
\end{equation}
%
with $\zeta$ being the chemical potential.
\par
%
In weak magnetic field, using the Euler-Maclaurin formula, the summation in $n$ in Eq.\ (\ref{eq:omega}) can be written as an integral in continuous variable $x$ and a residual term as
%
\begin{eqnarray}
\Omega \! &=& \! -\frac{1}{\beta} \frac{g_vg_s}{4\pi^2\hbar^2v^2} \! \sum_s \bigg( \! \int_{0}^\infty \!\! \varphi[\varepsilon_s(x)] dx \nonumber\\
&& \qquad - \frac{(\hbar\omega_B)^4}{12} \frac{\partial \varphi[\varepsilon_s(x)]}{\partial x} \Big|_{x=0} \bigg) + O(\delta^3).
\label{eq:omega_2}
\end{eqnarray}
%
The magnetization is given by
%
\begin{equation}
M = - \Big(\frac{\partial \Omega}{\partial B}\Big)_\zeta,
\end{equation}
%
and the magnetic susceptibility by
%
\begin{equation}
\chi = \lim_{B\rightarrow0} {M \over B} = -\Big(\frac{\partial^2 \Omega}{\partial B^2}\Big)_\zeta \Big|_{B=0}.
\label{eq:chi_def}
\end{equation}
%
\par
%
In Eq.\ (\ref{eq:omega_2}), the first term represents the thermodynamic potential in the absence of a magnetic field and only the second term depends on the magnetic field.
We have
%
\begin{eqnarray}
\chi = \int d\varepsilon \Big( -\frac{\partial f}{\partial \varepsilon} \Big) \chi(\varepsilon),
\end{eqnarray}
%
with
%
\begin{eqnarray}
\chi(\varepsilon) = -g_vg_s \frac{e^2 v^2}{6\pi c^2} \frac{1}{2|\Delta|} \theta(|\Delta|-|\varepsilon|),
\label{eq:chi_mono}
\end{eqnarray}
%
where $f(\varepsilon)$ is the Fermi distribution function.
The susceptibility at zero temperature is given by $\chi(\varepsilon_F)$ with $\varepsilon_F$ being the Fermi energy.
In the limit of $\Delta \to 0$, the susceptibility approaches a delta function
%
\begin{equation}
\chi(\varepsilon) = -g_vg_s \frac{e^2 v^2}{6\pi c^2} \delta(\varepsilon),
\end{equation}
%
in agreement with the previous result.\cite{McClure_1956a,Safran_and_DiSalvo_1979a,Koshino_and_Ando_2007b,Ando_2007d}
\par
\begin{figure}
\begin{center}
\leavevmode\includegraphics[width=1.\hsize]{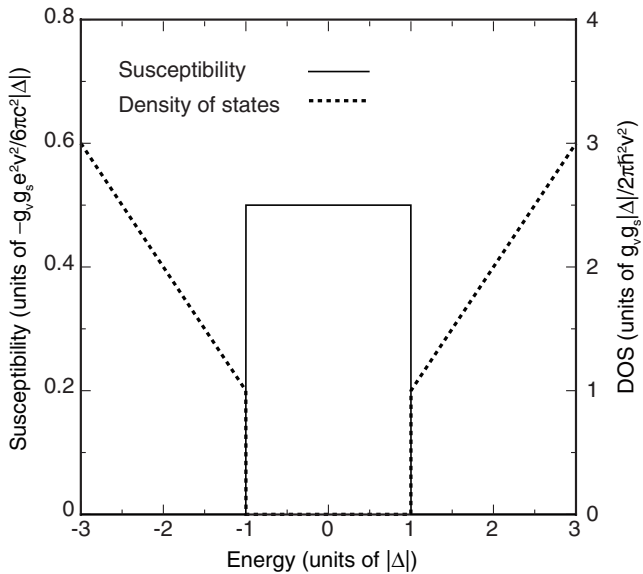}
\end{center}
\caption{Orbital susceptibility (solid) and density of states (dashed) of monolayer graphene with band gap $\Delta$.
Note that the upward direction represents negative (i.e., diamagnetic) susceptibility.
}
\label{fig_chi_mono}
\end{figure}

%
The susceptibility and the density of states, given by Eqs.\ (\ref{eq:chi_mono}) and (\ref{eq:DensityOfStatesInMonolayerGraphene}), respectively, are shown in Fig.\ \ref{fig_chi_mono}.
Note that the upward direction represents negative (i.e., diamagnetic) susceptibility. 
The susceptibility is not zero at zero electron density, $-1<\varepsilon/|\Delta|<+1$, because the completely filled valence band gives a constant diamagnetic susceptibility.
When the Fermi energy enters the conduction band, the susceptibility jumps downs to zero, resulting in zero total magnetism.
In the limit of $\Delta \to 0$, the step height at the band edge increases in proportion to $\Delta^{-1}$ and the susceptibility approaches a delta function.
\par
%
%
Because the Hamiltonian is equivalent to that of a Dirac electron with a nonzero mass, the magnetic susceptibility around the band edge should correspond to that of a conventional electron.
This is clearly illustrated by the effective Hamiltonian expanded in the vicinity of ${\bf k}=0$.
For the conduction band, $s=+1$, the effective Hamiltonian for the A site near the band bottom $(\varepsilon = \Delta)$ is written apart from the constant energy as
%
\begin{eqnarray}
\mathcal{H}^K &\approx& \frac{v^2}{2\Delta}\pi_-\pi_+ = \frac{\GVec{\pi}^2}{2m^*} - \frac{1}{2} g^* \mu_B B , \label{eq:kp_monolayer-K}\\
\mathcal{H}^{K'} &\approx& \frac{v^2}{2\Delta}\pi_+\pi_- = \frac{\GVec{\pi}^2}{2m^*} + \frac{1}{2} g^* \mu_B B, \label{eq:kp_monolayer-K'}
\end{eqnarray}
%
where $\mu_B$ is the Bohr magneton, given by $e\hbar/(2mc)$ with $m$ being the free electron mass, and we used the relation $[\pi_x,\pi_y] = i\hbar e B/c$ and defined
%
\begin{equation}
m^* = \frac{\Delta}{v^2}, \quad g^* =2 {m \over m^*} . 
\end{equation}
%
The last term in each Hamiltonian can be regarded as the pseudo-spin Zeeman term, where the different valleys $K$ and $K'$ serve as pseudo-spin up ($\xi=+1$) and down ($\xi=-1$), respectively. 
This agrees with the Zeeman energy expected for an intrinsic magnetic moment, that originates from the self-rotation of the wave packet in Bloch electron. \cite{Chang_and_Niu_1996,Xiao_et_al_2007}
The combined Hamiltonian is written as
%
\begin{equation}
\mathcal{H} \approx \frac{\GVec{\pi}^2}{2m^*} - \frac{\xi}{2} g^* \mu_B B .
\label{eq:H_conv}
\end{equation}
%
\par
%
Obviously, the pseudo-spin Zeeman term gives the Pauli paramagnetism and the first term containing $\GVec{\pi}^2$ gives the Landau diamagnetism in the usual form as
%
\begin{eqnarray}
\chi_P(\varepsilon) &=& \Big(\frac{g^*}{2}\Big)^2 \mu_B^2 D(\varepsilon) , \label{eq:chi_conv-P}\\
\chi_L(\varepsilon) &=& - {1\over 3 } \Big( \frac{m}{m^*}\Big)^2 \mu_B^2 D(\varepsilon), \label{eq:chi_conv-L}
\end{eqnarray}
%
with density of states
%
\begin{equation}
D(\varepsilon) = \frac{g_v g_s m^*}{2\pi\hbar^2} \, \theta(\varepsilon).
\end{equation}
%
The total susceptibility $\chi_P + \chi_L$ actually agrees with the amount of the jump at the conduction band bottom in $\chi$ of Eq.\ (\ref{eq:chi_mono}).
Because $g=2m/m^*$ in the present case, we have $\chi_L = -\chi_P/3 \propto 1/m^*$ as in the free electron, giving the paramagnetic susceptibility in total.
Therefore the susceptibility exhibits a discrete jump toward the paramagnetic direction when the Fermi energy moves off the Dirac point. 
The jump height goes to infinity as the gap closes, because the susceptibility is inversely proportional to the effective mass.
\par
%
\begin{figure}
\begin{center}
\leavevmode\includegraphics[width=0.9\hsize]{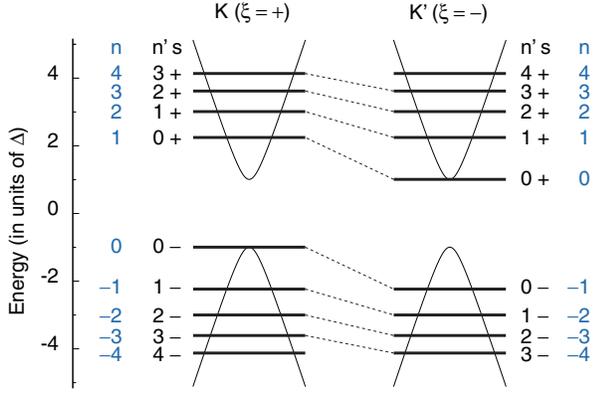}
\end{center}
\caption{(color online)
Landau-level energies of gapped monolayer graphene with for $\hbar\omega_B = 2\Delta$.  
Dashed lines connecting the levels of $K$ and $K'$ represent corresponding levels with opposite pseudo-spins.
}
\label{fig_mono_LL}
\end{figure}
%
In the original Hamiltonian, the Landau-level energies in Eq.\ (\ref{eq:mono_landauK}) can be rewritten as
%
\begin{eqnarray}
\varepsilon_{\xi,s,n'} = s \sqrt{(\hbar\omega_B)^2 \Big(n' + \frac 1 2 + \frac {\xi s}{2} \Big) + \Delta^2} \nonumber\\
(n' = 0,1,2,\cdots) .
\label{eq:mono_landau_uni}
\end{eqnarray}
%
Figure \ref{fig_mono_LL} shows energy levels for $\hbar\omega_B = 2\Delta$ and the relationship between the different labeling schemes of Eqs.\ (\ref{eq:mono_landauK}) and (\ref{eq:mono_landau_uni}).
For the conduction band, the levels of the same $n'$ with opposite pseudo-spins $\xi = \pm1$ share the same Landau level function $\phi_{n'}$  on the $A$ site, on which the states near the conduction-band bottom ($\varepsilon = \Delta$) have most of the amplitude as has been discussed above.
For the valence band, similarly, $n'$ describes the index of the Landau-level function at the $B$ site.
\par
%
\section{Bilayer graphene}
\label{sec_bi}
%
Bilayer graphene is a pair of graphene layers arranged in AB (Bernal) stacking and includes $A_1$ and $B_1$ atoms on layer 1 and $A_2$ and $B_2$ on layer 2.\cite{McCann_and_Falko_2006a,Guinea_et_al_2006a,Lu_et_al_2006a,Lu_et_al_2006b,McCann_2006a,Koshino_and_Ando_2006a,Nilsson_et_al_2006b,Partoens_and_Peeters_2006a,Partoens_and_Peeters_2007a}
The low energy states are again given by the states around $K$ and $K'$ points in the Brillouin zone.
The Hamiltonian at the $K$ point for the basis $(|A_1\rangle,|B_1\rangle$, $|A_2\rangle,|B_2\rangle)$ is given by
%
\begin{eqnarray}
{\cal H}^K =
\begin{pmatrix} \Delta & v \pi_- & 0 & 0 \\
v \pi_+  & \Delta & \gamma_1 & 0 \\
0  & \gamma_1 & -\Delta &  v \pi_- \\
0 & 0 &  v \pi_+ & -\Delta
\end{pmatrix} ,
\label{eq:H_bi}
\end{eqnarray}
%
where $\Delta$ describes potential difference between layer 1 and 2 (not $A$ and $B$ sites) and $\gamma_1$ represents interlayer coupling between $B_1$ and $A_2$.\cite{McCann_and_Falko_2006a,Guinea_et_al_2006a,Ando_and_Koshino_2009a}
The Hamiltonian at the $K'$ point is obtained by exchanging $\pi_\pm$ in Eq.\ (\ref{eq:H_bi}).
\par
\begin{figure}
\begin{center}
\includegraphics[width=0.90\hsize]{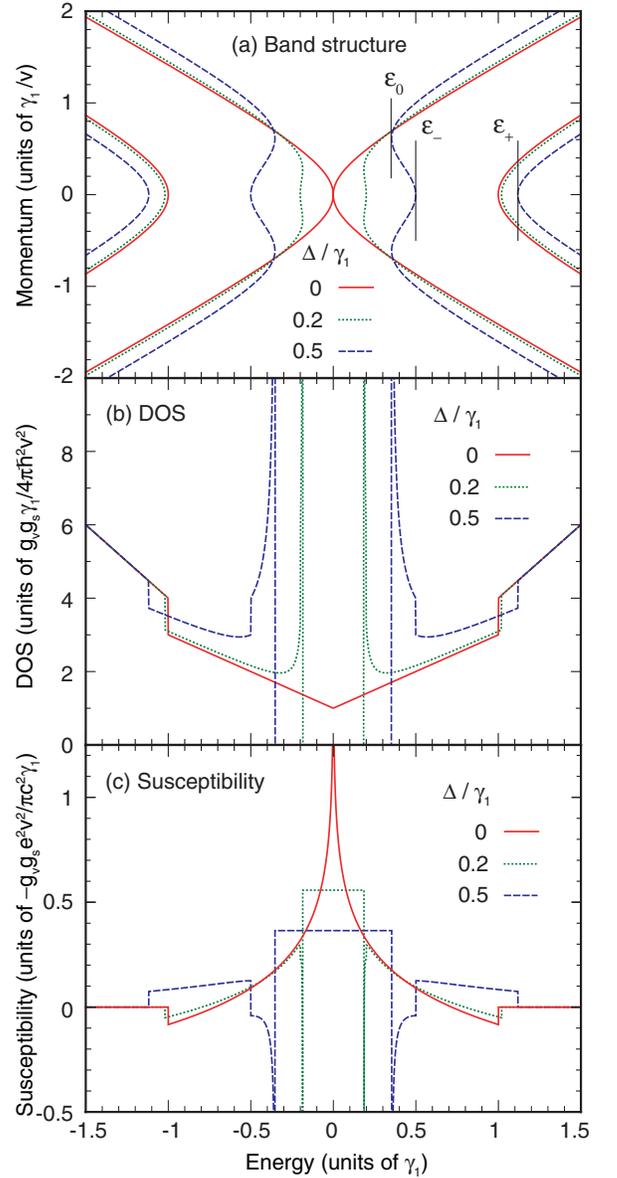}
\caption{(color online)
(a) Band structure, (b) density of states, and (c) susceptibility of bilayer graphenes with the asymmetry gap $\Delta/\gamma_1 = 0$, 0.2, and 0.5.
Vertical lines in (a) indicate the energies of $\varepsilon_0$, $\varepsilon_-$ and $\varepsilon_+$ for $\Delta/\gamma_1 = 0.5$.
The upward direction represents negative (i.e., diamagnetic) susceptibility in (c).
}
\label{fig_chi_bilayer}
\end{center}
\end{figure}

%
The energy band at $B=0$ is given by
%
\begin{eqnarray}
\varepsilon_{s\mu}(p) \!\! &=& \!\! s \Big( \frac{\gamma_1^2}{2} + v^2 p^2 + \Delta^2 \nonumber \nonumber\\
&& + \mu \Big[ \frac{\gamma_1^4}{4} + v^2p^2(\gamma_1^2 + 4\Delta^2) \Big]^{1/2} \Big)^{1/2} ,
\label{eq:bilayer_band}
\end{eqnarray}
%
with $\mu = \pm1$.\cite{McCann_2006a}
The index $\mu= +1$ and $-1$ give a pair of bands further and closer to zero energy, respectively, and $s=+1$ and $-1$ in each pair represent the electron and hole branches, respectively.
The band-edge energies corresponding to $p=0$ are given by $|\varepsilon| = \varepsilon_\pm$ for $\mu=\pm1$, where 
%
\begin{equation}
\varepsilon_+ = \sqrt{\gamma_1^2+\Delta^2}, \quad
\varepsilon_- = |\Delta|.
\end{equation}
%
For $\mu =-1$, the band minimum becomes
%
\begin{equation}
\varepsilon_0 = \frac{\gamma_1|\Delta|}{\sqrt{\gamma_1^2+4\Delta^2}},
\end{equation}
%
which corresponds to an off-center momentum.\cite{Guinea_et_al_2006a}
The density of states diverges here as $D(\varepsilon) \propto (\varepsilon-\varepsilon_0)^{-1/2}$.
The energy bands and the density of states with several $\Delta$'s are plotted in Figs.\ \ref{fig_chi_bilayer} (a) and (b), respectively.
Vertical lines in (a) indicate the energies of $\varepsilon_0$, $\varepsilon_-$, and $\varepsilon_+$ for $\Delta = 0.5\gamma_1$.
\par
%
%
In a magnetic field, the eigenfunction of the Hamiltonian at the $K$ point is written as $(c_1 \phi_{n-1}, \allowbreak c_2 \phi_n, \allowbreak c_3 \phi_n, \allowbreak c_4\phi_{n+1})$ with integer $n\geq -1$.
For $n \geq 1$, the Hamiltonian matrix for $(c_1,c_2,c_3,c_4)$ becomes\cite{Guinea_et_al_2006a, Koshino_and_McCann_2009c}
%
\begin{equation}
H^K_{n\geq 1} = \begin{pmatrix}
\Delta & \hbar\omega_B\sqrt{n} & 0 & 0 \\
\hbar\omega_B\sqrt{n}  & \Delta & \gamma_1 & 0 \\
0  & \gamma_1 & -\Delta &  \hbar\omega_B\sqrt{n+1} \\
0 & 0 &  \hbar\omega_B\sqrt{n+1} & -\Delta
\end{pmatrix} ,
\label{eq:bilayer_matrix_k}
\end{equation}
%
For $n=0$, the first component does not actually exist because $\phi_{-1} = 0$.
The matrix for $(c_2,c_3,c_4)$ becomes
%
\begin{equation}
H^K_{0} = \begin{pmatrix}
\Delta & \gamma_1 & 0 \\
\gamma_1 & -\Delta &  \hbar\omega_B \\
0 &  \hbar\omega_B & -\Delta
\end{pmatrix} .
\end{equation}
%
For $n=-1$, only the component $c_4$ survives and the Hamiltonian is
%
\begin{equation}
H^K_{-1} = -\Delta.
\end{equation}
%
\par
%
For the $K'$ point, the eigenfunction is written as $(c_1 \phi_{n+1}, \allowbreak c_2 \phi_n, \allowbreak c_3 \phi_n, \allowbreak c_4\phi_{n-1})$.
For $n \geq 1$, the Hamiltonian matrix for $(c_1,c_2,c_3,c_4)$ is
%
\begin{equation}
H^{K'}_{n\geq 1} = \begin{pmatrix}
\Delta & \hbar\omega_B\sqrt{n+1} & 0 & 0 \\
\hbar\omega_B\sqrt{n+1}  & \Delta & \gamma_1 & 0 \\
0  & \gamma_1 & -\Delta &  \hbar\omega_B\sqrt{n} \\
0 & 0 &  \hbar\omega_B\sqrt{n} & -\Delta
\end{pmatrix} .
\label{eq:bilayer_matrix_k2}
\end{equation}
%
For $n=0$, the matrix for $(c_1,c_2,c_3)$ becomes
%
\begin{equation}
H^{K'}_{0} = \begin{pmatrix}
\Delta & \hbar\omega_B & 0 \\
\hbar\omega_B  & \Delta &  \gamma_1 \\
0 &  \gamma_1 & -\Delta
\end{pmatrix} ,
\end{equation}
%
and for $n=-1$, that for $c_1$ is
%
\begin{equation}
H^{K'}_{-1} = \Delta.
\end{equation}
%
\par
%
\begin{figure}
\begin{center}
\includegraphics[width=0.9\hsize]{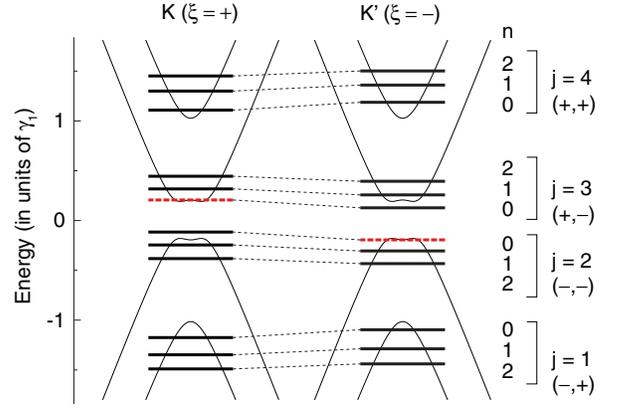}
\caption{(color online) 
Landau-level energies of bilayer graphene given by Eq.\ (\ref{eq:bilayer_matrix}) with $\Delta = 0.2\gamma_1$ and $\hbar\omega_B = 0.5\gamma_1$.
The dashed horizontal lines represent the Landau level which originally belongs to $n=-1$ at opposite valley.
The quantum number $(s,\mu)$ is indicated below $j$.}
\label{fig_bi_LL}
\end{center}
\end{figure}
%
If we extend the definition of the matrix of Eq.\ (\ref{eq:bilayer_matrix_k}) to $n=0$, its three eigenvalues agree with those of $H^{K}_0$ and the rest with that of $H^{K'}_{-1}$.
Similarly, the matrix of Eq.\ (\ref{eq:bilayer_matrix_k2}) with $n=0$ gives eigenvalues of $H^{K'}_0$ and $H^{K}_{-1}$.
Thus we can use Eqs.\ (\ref{eq:bilayer_matrix_k}) and (\ref{eq:bilayer_matrix_k2}) with $n\geq 0$ to produce the full spectrum.
By introducing the pseudo-spin variable $\xi = \pm1$, the Hamiltonian is combined into a single expression,
%
\begin{equation}
H^\xi_{n} = \begin{pmatrix}
\Delta & \sqrt{x_{n-}} & 0 & 0 \\
\sqrt{x_{n-}}  & \Delta & \gamma_1 & 0 \\
0  & \gamma_1 & -\Delta &  \sqrt{x_{n+}} \\
0 & 0 & \sqrt{x_{n+}} & -\Delta
\end{pmatrix} ,
\label{eq:bilayer_matrix}
\end{equation}
%
with
%
\begin{eqnarray}
x_{n\pm} = x_n \pm \frac{1}{2} \xi \delta,
\end{eqnarray}
%
and
%
\begin{equation}
x_n = \left(n + \frac{1}{2}\right) \delta, \quad \delta =  (\hbar\omega_B)^2.
\label{eq:xn}
\end{equation}
%
We write the eigenvalues of $H^\xi_{n}$ as
%
\begin{equation}
 \varepsilon_{j}(x_n, \xi \delta) \quad (j=1,2,3,4),
\label{eq:eigen}
\end{equation}
%
in the ascending order in energy ($j=1$ and 2 for valence bands and $j=3$ and 4 for the conduction bands).
The second argument in $\varepsilon_{j}(x_n, \xi \delta)$ represents the dependence on $B$ which are not included in $x_n$.
\par
%
Figure \ref{fig_bi_LL} shows the example of the Landau-level spectrum at $\Delta/\gamma_1 = 0.2$ and $\hbar\omega_B/\gamma_1 = 0.5$, where the thick dashed lines represent the Landau level which originally belongs to $n=-1$ at opposite valleys.
The correspondence between quantum numbers $j$ and $(s,\mu)$ are indicated in the figure.
\par
%
The thermodynamic potential becomes
%
\begin{eqnarray}
\Omega &=& -\frac{1}{\beta} \frac{g_s}{2\pi l_B^2} \sum_{\xi, j} 
\sum_{n=0}^\infty 
\varphi\big[\varepsilon_{j}(x_n,\xi\delta) \big]
\nonumber\\
&=&
 -\frac{1}{\beta} \frac{g_s}{4\pi \hbar^2 v^2}\! \sum_{\xi, j} 
\Bigg[ \! \int_{0}^\infty \!\!
\varphi\big[\varepsilon_{j}(x,\xi\delta) \big]  dx 
\nonumber\\
&& + \frac{\delta^2}{24} 
\frac{\partial \varphi[\varepsilon_{j}(x,0)]}{\partial x}
\Big|_{x=0}
 \Bigg] 
+ O(\delta^3),
\label{eq:omega_bilayer}
\end{eqnarray}
%
where we used the Euler-Maclaurin formula in the second equation.
The first term in the bracket can be transformed by changing the integral variable from $x$ to $\varepsilon$ as
%
\begin{equation}
\frac{1}{\beta} \int_{0}^\infty \!\!
\varphi\big[\varepsilon_{j}(x,\xi\delta) \big]  dx 
= \int_{-\infty}^\infty \!\!
f(\varepsilon) \, n_{j}(\varepsilon,\xi\delta)
d\varepsilon,
\end{equation}
%
where we used $\varphi'(\varepsilon) = -\beta f(\varepsilon)$ and defined
%
\begin{equation}
n_{j}(\varepsilon, \xi\delta) \equiv
s_{j}(\varepsilon, \xi\delta) x_{j}(\varepsilon, \xi\delta),
\end{equation}
%
where
%
$x_{j}(\varepsilon, \xi\delta)$ is a real and positive solution of $\varepsilon = \varepsilon_{j}(x,\xi\delta)$ and
%
\begin{equation}
s_{j}(\varepsilon, \xi\delta) \equiv \textrm{sgn} \Big(\frac{\partial x_{j}(\varepsilon,\xi\delta)}{\partial \varepsilon} \Big) .
\end{equation}
%
If there are more than one solution of $x_{j}$, we regard $n_{j}$ as their sum.
The quantity $n_{j}(\varepsilon,\xi\delta)/(4\pi \hbar^2 v^2)$ represents the electron density below $\varepsilon$ for the conduction band and the hole density above $\varepsilon$ for the valence band.
\par
%
By expanding
%
\begin{equation}
n_{j}(\varepsilon, \xi\delta) = n^{(0)}_{j}(\varepsilon) + n^{(1)}_{j}(\varepsilon) \, \xi \delta + \frac{1}{2} n^{(2)}_{j}(\varepsilon) \, \delta^2 + \cdots,
\label{eq:n_expand}
\end{equation}
%
we can further expand $\Omega$ of Eq.\ (\ref{eq:omega_bilayer}) in terms of $\delta \propto B$.
We have
%
\begin{eqnarray}
\chi(\varepsilon) \! &=& \! g_s g_v 
\frac{e^2 v^2}{\pi c^2}
\sum_{j}
\Biggl[
\int_{-\infty}^\varepsilon \!\! 
n_{j}^{(2)}(\varepsilon')
d \varepsilon'
\nonumber\\
&& \!\!\! 
- \frac{1}{12} 
\theta\bigl[\varepsilon-\varepsilon_{j}(0,0)\bigr]
\frac{\partial \varepsilon_{j}(x,0)}{\partial x} \Big|_{x=0}
\Biggr]. \quad
\label{eq:chi_form}
\end{eqnarray}
%
\par
%
For the Hamiltonian of Eq.\ (\ref{eq:bilayer_matrix}), the eigenequation $\det (\varepsilon -H^\xi_{n}) =0$ can be solved for $x$ $(\equiv x_n)$ as
%
\begin{equation}
x_\pm  = \varepsilon^2 + \Delta^2 \pm \frac{1}{2}
\sqrt{
(4\varepsilon \Delta - \xi\delta)^2 
+ 4\gamma_1^2(\varepsilon^2-\Delta^2)
} ,
\label{eq:x_sol}
\end{equation}
%
which gives $x_{j}(\varepsilon,\xi\delta)$ when being real and positive.
Let us first consider the case $\varepsilon>\varepsilon_+$, where two conduction bands are occupied by electrons.
In this case $x_\pm$ are both real and positive and we have $x_1=x_2=0$, $x_3=x_+$, and $x_4=x_-$.
Then, we have
%
\begin{equation}
\sum_{j} n_{j}(\varepsilon,\xi\delta) = x_+ + x_- =
2(\varepsilon^2+\Delta^2),
\label{eq:xsum}
\end{equation}
%
independent of $\xi\delta$.
Therefore, $\sum_jn^{(2)}_j(\varepsilon)$ identically vanishes, resulting in susceptibility independent of energy in the region $\varepsilon> \varepsilon_+$.
The same is true for $\varepsilon < -\varepsilon_+$.
Because $\chi = 0$ for $\varepsilon = \pm \infty$, i.e., in the case of empty or filled band, we can conclude that the susceptibility vanishes for $\varepsilon>\varepsilon_+$ and $\varepsilon<-\varepsilon_+$ independent of interlayer interaction $\gamma_1$ and asymmetry $\Delta$.
\par
%
Similarly, the density of states for $|\varepsilon|> \varepsilon_+$ is independent of $\gamma_1$ and $\Delta$ and becomes twice as large as that of monolayer.
In fact, we have
%
\begin{equation}
D(\varepsilon) \propto \frac{\partial}{\partial \varepsilon} \sum_j n_{j}(\varepsilon,0) = 4\varepsilon.
\end{equation}
%
This feature of the density of states is apparent in Fig.\ \ref{fig_chi_bilayer} (b).
\par
%
In the vicinity of the bottom of the excited conduction band, $\varepsilon = \varepsilon_+$, we have
%
\begin{eqnarray}
n^{(2)}_{4}(\varepsilon) &=& \frac{\partial^2}{\partial \delta^2} x_-\theta(x_-)\Bigr|_{\delta=0} \nonumber\\
&=& \Big[ \frac{\partial^2 x_-}{\partial \delta^2}\theta(x_-) + \Big( \frac{\partial x_-}{\partial \delta} \Big)^2 \delta(x_-) \Big]_{\delta=0},
\label{eq:n4}
\end{eqnarray}
%
where we used $x_-\delta(x_-) = 0$ and $x_-\delta'(x_-) = -\delta(x_-)$. 
Using Eq.\ (\ref{eq:chi_form}), we find that the susceptibility makes a discrete jump at $\varepsilon_+$ as
%
\begin{eqnarray}
&& \chi(\varepsilon_+ + 0) -  \chi(\varepsilon_+ -0)
\nonumber \\
&& = 
g_v g_s \frac{e^2 v^2}{\pi c^2}
\Big(
 \frac{\Delta^2\sqrt{\Delta^2+\gamma_1^2}}{\gamma_1^2(2\Delta^2 + \gamma_1^2)}
- \frac{2\Delta^2 + \gamma_1^2}{12\gamma_1^2\sqrt{\Delta^2+\gamma_1^2}}
\Big), \qquad
\label{eq:chi_bi_jump}
\end{eqnarray}
%
where the first term in the bracket comes from the integral of the delta function in Eq.\ (\ref{eq:n4}) and the second term from the step function in Eq.\ (\ref{eq:chi_form}).
\par
%
Near $\varepsilon_+$, the eigenstates are given primarily by the dimer states composed of $|B_1\rangle$ and $|A_2\rangle$.
The effective Hamiltonian is described by the second order in interband interation with the conduction-band bottom $|A_1\rangle$ and the valence-band top $|B_2\rangle$, where each process gives a term $\propto\pi_+\pi_-$ or $\propto\pi_-\pi_+$.
In symmetric bilayer with $\Delta = 0$, the terms $\pi_+\pi_-$ and $\pi_-\pi_+$ have the same coefficient and the pseudo-spin Zeeman term identically vanishes.
When $\Delta$ becomes nonzero, the two coefficients shift from each other linearly in $\Delta$ because of the band-gap opening, leading to a nonzero Zeeman term.
The resulting effective Hamiltonian is given by Eq.\ (\ref{eq:H_conv}) with
%
\begin{eqnarray}
 m^* = \frac{\gamma_1^2\sqrt{\Delta^2+\gamma_1^2}}
{2v^2(2\Delta^2+\gamma_1^2)},
\quad 
g^* = \frac{4\Delta\sqrt{\Delta^2+\gamma_1^2}}
{2\Delta^2+\gamma_1^2} \frac{m}{m^*}.
\label{eq:bi_g}
\end{eqnarray}
%
\par
%
The susceptibility is written as Pauli and Landau magnetism in Eqs.\ (\ref{eq:chi_conv-P}) and (\ref{eq:chi_conv-L}), respectively, which together give a susceptibility jump of Eq.\ (\ref{eq:chi_bi_jump}).
The paramagnetic component $\chi_P$ is zero at $\Delta = 0$ and monotonically increases as $\Delta$ becomes larger.
At $g^* = (2/\sqrt{3})(m/m^*)$ or $\Delta \approx 0.34 \gamma_1 $,  $\chi_P$ exceeds $\chi_L$ and the susceptibility step changes from diamagnetic to paramagnetic.
In the limit $\Delta \to \infty$, we have $g^*=2m/m^*$ as in the monolayer.
This is to be expected, because the bilayer graphene in this limit can be regarded as a pair of independent monolayer graphenes, where interlayer coupling $\gamma_1$ opens an energy gap at each Dirac point.
Similar argument also applies to the behavior around $\varepsilon_-$.
\par
%
In the energy region $-\varepsilon_- < \varepsilon < -\varepsilon_0$ near the top of the valence band, both $x_+$ and $x_-$ are real and positive, giving the states at outer and inner equi-energy circle of the band $j=2$, respectively.
Then we have
%
\begin{eqnarray}
n^{(2)}_2(\varepsilon) &=& 
\frac{\partial^2}{\partial \delta^2}
(-x_+ + x_-)\Bigr|_{\delta=0}
\nonumber\\
&=&
\frac{\gamma_1^2(\Delta^2 -\varepsilon^2)}
{2 [(4\Delta^2 + \gamma_1^2) 
(\varepsilon^2 - \varepsilon_0^2)]^{3/2}}.
\end{eqnarray}
%
When the energy approaches to $-\varepsilon_0$ from the negative side, the integral of $n^{(2)}_2(\varepsilon)$, thus the susceptibility, diverges in positive direction as $\propto (\varepsilon+\varepsilon_0)^{-1/2}$ in the same manner as the density of states.
The same divergence occurs at the bottom of the conduction band, $+\varepsilon_0$, because of the electron-hole symmetry.
\par
%
Full analytic expression of the susceptibility $\chi(\varepsilon)$ is complicated and presented in Appendix \ref{sec_app}.
Figure \ref{fig_chi_bilayer} (c) plots the susceptibility for $\Delta=0$, 0.2, and 0.5.
In accordance with the above analytic consideration, we actually observe that the susceptibility vanishes in the region $\varepsilon > \varepsilon_+$ and $\varepsilon < - \varepsilon_+$ and that the susceptibility step at $\varepsilon = \varepsilon_+$ changes from diamagnetic to paramagnetic with increasing $\Delta$.
We also see that the susceptibility for $\Delta\neq 0$ diverges in the paramagnetic direction at $\varepsilon = \pm \varepsilon_0$.
\par
%
\section{Three dimensional Dirac system}
\label{sec_3d}
%
The results in monolayer graphene in Sec.\ \ref{sec_mono} can be directly extended to three-dimensional Dirac Hamiltonian, which is also known to describe the approximate electronic structure of bismuth with strong spin-orbit interaction.\cite{Wolff_1964a,Fukuyama_and_Kubo_1969a,Fukuyama_and_Kubo_1970a,Fuseya_et_al_2009a}
In bisumuth, electronic states near the Fermi level is approximately described by a $(4,4)$ matrix Hamiltonian, given by
%
\begin{equation}
\mathcal{H} =
\begin{pmatrix}
 \Delta & 0 &  v \pi_z &  v \pi_- \\
 0 & \Delta &  v \pi_+ & -  v \pi_z \\
 v \pi_z &  v \pi_- & -\Delta & 0 \\
 v \pi_+ & - v \pi_z & 0 & -\Delta
\end{pmatrix} ,
\end{equation}
%
where four components consist of two orbital and two spin degerees of freedom.
The density of states at zero magnetic field is 
%
\begin{eqnarray}
 D(\varepsilon) = \frac{g_vg_s}{\pi^2\hbar^3 v^3} |\varepsilon|\sqrt{\varepsilon^2-\Delta^2}
\,\theta(\varepsilon^2-\Delta^2),
\end{eqnarray}
%
where $g_v$ is the valley degeneracy allowing the presence of different $k$ points described by the above Hamiltonian in the first Brillouin zone.
\par
%
The Landau levels in a uniform magnetic field in $z$ direction are given by
%
\begin{eqnarray}
\varepsilon_{s, n, \sigma}
&=& s \sqrt{(\hbar \omega_B)^2 
\Big( n +\frac 1 2 + \frac \sigma 2 \Big)  + v^2 p_z^2 + \Delta^2} \nonumber\\
&& \hspace{30mm} (n=0,1,2,\cdots),
\label{eq:3d_landau}
\end{eqnarray}
%
with $\hbar\omega_B = \sqrt{2}\hbar v/l_B$, $s=\pm1$, and $\sigma = \pm1$.
This is equivalent to the two-dimensional Dirac system, Eq.\ (\ref{eq:mono_landau_uni}), when the term $\Delta^2$ is replaced with $\Delta^2+v^2 p_z^2$.
The susceptibility $\chi(\varepsilon)$ is thus calculated by integrating Eq.\ (\ref{eq:chi_mono}) in $p_z$ as
%
\begin{eqnarray}
\chi(\varepsilon) \!\! &=& \!\! -\frac{g_vg_s e^2 v^2}{6\pi c^2}  \int \frac{dp_z}{2\pi \hbar} \frac{\theta(\Delta^2+v^2 p_z^2-\varepsilon^2)}{2\sqrt{\Delta^2+v^2 p_z^2}} \nonumber \\
&=& \!\! -\frac{g_vg_s e^2 v}{12\pi^2\hbar c^2}
\left\{
\begin{array}{ll}
\displaystyle \log \frac{2\varepsilon_c}{|\Delta|} & (|\varepsilon|<|\Delta|); \\
\displaystyle \log \frac{2\varepsilon_c}{|\varepsilon| + \sqrt{\varepsilon^2 - \Delta^2}} & (|\varepsilon|>|\Delta|),
\end{array}
\right.
\qquad
\end{eqnarray}
%
where $\varepsilon_c$ is a cut-off energy.
In the limit of $\Delta \to 0$, the susceptibility at zero energy logarithmically diverges.
\par
%
\begin{figure}
\begin{center}
\includegraphics[width=1.\hsize]{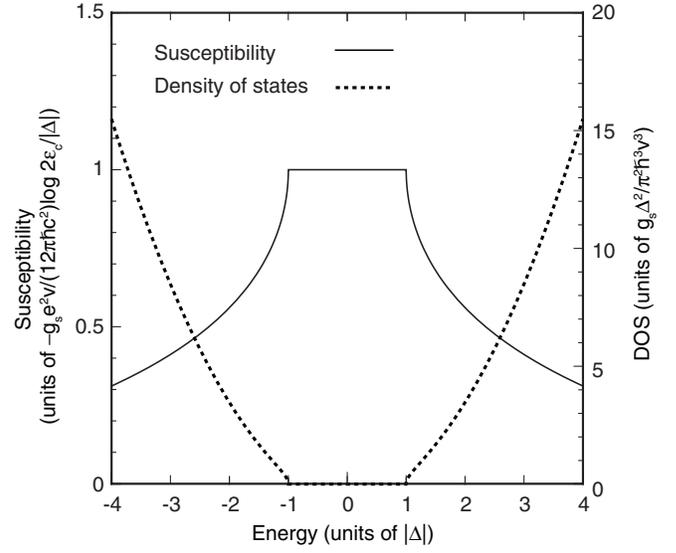}
\end{center}
\caption{Orbital susceptibility and density of states of three-dimensional Dirac electron.}
\label{fg_chi_3d_dirac}
\end{figure}
%
At an energy $\varepsilon$ just above the band bottom $|\Delta|$, we obtain the paramagnetic contribution
%
\begin{eqnarray}
\chi(\varepsilon) - \chi(0) \approx \frac{2}{3} \Big({m \over m^*} \Big)^2 D(\varepsilon) \mu_B^2 ,
\end{eqnarray}
%
where $D(\varepsilon)=(g_sg_v/4\pi^2)(2m^*/\hbar^2)^{3/2}\sqrt{\varepsilon}$ with $m^* = \Delta/v^2$.
This is nothing but the magnetic susceptibility, dominated by the Pauli paramagnetism, of a three-dimensional metal with mass $m^*$ and $g$ factor $g^*=2m/m^*$.
Figure \ref{fg_chi_3d_dirac} shows the susceptibility and the density of states in the present system.
The singular decrease of the susceptibility at the band edges is fully understood in terms of the appearance of the dominant spin paramagnetism inside the band.
\par
%
\section{Conclusion}
\label{sec_conc}
%
We have shown that the orbital magnetism singularly dependent on the Fermi level, appearing in narrow gap electronic systems described by the Dirac Hamiltonian, can be understood in terms of pseudo-spin Pauli paramagnetism induced by extra degree of freedom in the orbital motion.
This has been demonstrated by explicit calculations of orbital susceptibility in monolayer and bilayer graphenes with band gap and also in three dimensional Dirac systems such as bismuth.
\par
%
\section*{ACKNOWLEDGMENTS}
%
This work was supported in part by Grant-in-Aid for Scientific Research on Priority Area ``Carbon Nanotube Nanoelectronics,'' by Grant-in-Aid for Scientific Research, and by Global Center of Excellence Program at Tokyo Tech ``Nanoscience and Quantum Physics'' from Ministry of Education, Culture, Sports, Science and Technology Japan.
\par
%
\appendix
%
\section{Susceptibility of bilayer graphene}
\label{sec_app}
%
Using Eqs.\ (\ref{eq:chi_form}) and (\ref{eq:x_sol}), the susceptibility of $\chi(\varepsilon)$ of bilayer graphene with energy gap is calculated as
%
\begin{eqnarray}
\chi(\varepsilon) = g_v g_s \frac{e^2 v^2}{\pi c^2 \gamma_1}
\tilde\chi(\varepsilon),
\end{eqnarray}
%
with 
%
\begin{eqnarray}
\tilde\chi(\varepsilon) =
\left\{
\begin{array}{ll}
\displaystyle \tilde\chi_0 + \tilde\chi_- + \tilde\chi_+ & (|\varepsilon| < \varepsilon_0), \\
2F(\varepsilon) + \tilde\chi_- + \tilde\chi_+ & (\varepsilon_0 < |\varepsilon| < \varepsilon_-), \\
F(\varepsilon) + \tilde\chi_+ & (\varepsilon_- < |\varepsilon| < \varepsilon_+), \\
0 & (\varepsilon_+ < |\varepsilon|).
\end{array}
\right.
\end{eqnarray}
%
where
%
\begin{eqnarray}
F(\varepsilon) \!\! &=& \!\! \frac{\gamma_1 \Delta ^2 |\varepsilon| } {(\gamma _1^2+4 \Delta^2 ) \sqrt{\gamma _1^2 (\varepsilon ^2-\Delta^2 ) + 4 \Delta ^2 \varepsilon ^2}} \nonumber\\
&& + \frac{\gamma _1^3}{4 \left(\gamma _1^2+4 \Delta^2\right){}^{3/2}} \log\Big[2|\varepsilon| (\gamma _1^2 +4 \Delta ^2) \nonumber\\
&& + 2 \sqrt{\gamma _1^2+4 \Delta ^2} \sqrt{\gamma_1^2 \left(\varepsilon ^2-\Delta ^2\right)+4 \Delta ^2 \varepsilon^2} \Big] , \quad
\end{eqnarray}
%
and
%
\begin{eqnarray}
\tilde\chi_0 \!\! &=& \!\! \frac{\gamma_1^3\log[4\gamma_1^2\Delta^2(\gamma_1^2+4\Delta^2)]} {4(\gamma_1^2+4\Delta^2)^{3/2}} , \\
\tilde\chi_- \!\! &=& \!\! -F(\varepsilon_-) + \frac{|\Delta|}{3\gamma_1}, \\
\tilde\chi_+ \!\! &=& \!\! -F(\varepsilon_+) - \frac{\Delta^2\sqrt{\Delta^2+\gamma_1^2}}{\gamma_1(2\Delta^2 + \gamma_1^2)} + \frac{2\Delta^2 + \gamma_1^2}{12\gamma_1\sqrt{\Delta^2+\gamma_1^2}}. \qquad 
\end{eqnarray}
%
In the symmetric bilayer with vanishing gap, $\Delta = 0$, in particular, we simply get
%
\begin{eqnarray}
\chi(\varepsilon) = g_v g_s \frac{e^2 v^2}{\pi c^2 \gamma_1}
 \theta(\gamma_1-|\varepsilon|)
\Big(
\frac{1}{4} \log \frac{|\varepsilon|}{\gamma_1} + \frac{1}{12}
\Big) ,
\label{eq:chi_bi_d0}
\end{eqnarray}
%
which agrees with the previous results.\cite{Safran_1984a,Koshino_and_Ando_2007c} 
\par
%

%
\vspace{0.150cm}
\hrule
\vspace{0.150cm}
\rightline{File: \jobname.tex (\today)}
%
\end{document}